\documentclass[12pt]{article}

\usepackage{graphics,psfrag}
\usepackage{amsthm,amssymb,epsfig,amsmath,euscript,array,cite}
\usepackage{hyperref}

\newcommand{\be}{\begin{equation}}
\newcommand{\ee}{\end{equation}}
\def\bea{\begin{eqnarray}}
\def\eea{\end{eqnarray}}
\newcommand{\eq}[1]{(\ref{#1})}
\def\nn{\nonumber}
\def\del{\partial}

\newcommand{\beq}{\begin{equation}}
\newcommand{\eeq}{\end{equation}}
\newcommand{\ben}{\begin{eqnarray}}
\newcommand{\een}{\end{eqnarray}}
\newcommand{\bes}{\begin{subequations}}
\newcommand{\ees}{\end{subequations}}
\newcommand{\blg}{\begin{align}}
\newcommand{\elg}{\end{align}}

\newcommand{\cN}{{\cal N}}

\newcommand{\Ncal}{\mathcal{N}}

\newcommand{\AdS}{\mathrm{AdS}}

\newcommand{\prt}[1]{{\left( {#1} \right)}}
\newcommand{\la}[1]{\label{#1}}

\def\one{\mbox{1 \kern-.59em {\rm l}}}

\def\da{{\dot\alpha}}
\def\dth{{\dot\theta}}
\def\dphi1{{\dot\phi_1}}
\def\dphi2{{\dot\phi_2}}
\def\dphi3{{\dot\phi_3}}
\def\dphi{{\dot\phi}}

\def\a{\alpha}      \def\da{{\dot\alpha}}
\def\b{\beta}

\def\k{\kappa}

\def\s{\sigma}  
\def\t{\tau}
\def\th{\theta}

 \def\cN{{\cal N}}

\def\gtt{\tilde{\gamma}}
\def\stt{\tilde{\sigma}}
\newcommand{\comment}[1]{}

\newcommand{\diff}{{\mathrm d}}

\newcommand{\p}{\partial}

\newcommand{\Hcal}{\mathcal{H}}

\begin{document}

\hfill MCTP-13-34

\vspace{15pt}

\begin{center}

{\Large \bf
On Marginal Deformations and Non-Integrability
}
\vspace{20pt}

\vspace{20pt}
{\bf
Dimitrios Giataganas$^{a,b}$, Leopoldo A. Pando Zayas$^c$ and Konstantinos Zoubos$^d$\hspace{-.92cm}
}

{\em
$^a$Physics Division, National Technical University of Athens\\
15780 Zografou Campus, Athens, Greece
}

{\em
$^b$Department of Physics, University of Athens\\
15771 Athens, Greece
}

{\em
$^c$Michigan Center for Theoretical Physics, University of Michigan\\
Ann Arbor, MI 48109, USA}

{\em
$^d$Department of Physics, University of Pretoria\\
Private Bag X20, Hatfield 0028, South Africa}

{\small \sffamily
dimitrios.giataganas@gmail.com, lpandoz@umich.edu, kzoubos@up.ac.za
}
\end{center}

\vspace{20pt}
\begin{center} {\bf Abstract}\end{center}

We study the interplay between a particular marginal deformation of $\Ncal=4$ super Yang-Mills theory, the $\beta$ deformation, and integrability in the holographic setting. Using modern methods of analytic non-integrability of Hamiltonian systems, we find that, when the $\beta$ parameter takes imaginary values, classical string trajectories on the dual background become non-integrable. We expect the same to be true for generic complex $\beta$ parameter. By exhibiting the Poincar\'e sections and phase space trajectories for the generic complex $\beta$ case, we provide numerical evidence of strong sensitivity to initial conditions. Our findings agree with expectations from weak coupling that the complex $\beta$ deformation is non-integrable and provide a rigorous argument beyond the trial and error approach to non-integrability.

\pagebreak

%%%%%%%%%%%%%%%%%%%%%%%%%%%%%%%%%%%%%%%%%%%%%%%%%%%%%%%%%%%%%%%%%%%%%%%%%%%%%%%
\section{Introduction}

The AdS/CFT correspondence is a powerful tool that has provided a bridge connecting field theory and gravity \cite{Maldacena:1997re,Witten:1998qj,Gubser:1998bc}. In its most powerful setting it implies a mathematical identification between strings in AdS${}_5 \times S^5$ with Ramond-Ramond flux and ${\cal N}=4$ supersymmetric Yang-Mills theory (sYM). It is particularly insightful in the strong coupling limit of field theories.

Integrability plays a fundamental role in the AdS/CFT correspondence, and has brought us closest to the potential solution of $\Ncal=4$ sYM  in the planar limit \cite{Beisert:2010jr}. In the planer limit of ${\cal N}=4$ sYM and in the dual string theory on AdS${}_5\times S^5$ integrability is beyond doubt \cite{Beisert:2010jr}. The advances achieved using integrability naturally beg the question of whether these techniques can be applied in a wider context. For example, beyond the planar limit of $\Ncal=4$ sYM, or to the deformations of sYM. It is, therefore, particularly instructive to bring the topic of non-integrability to weigh on in particular modifications of  ${\cal N}=4$ sYM to place a bound on the power of integrability in the wider context.

To motivate the role of non-integrability from a more general perspective, we first recall that some of the key advances in our understanding of the AdS/CFT correspondence have been propelled by semiclassical quantization. Semiclassical quantization is the best way to approach backgrounds with RR charges where standard techniques in string theory fail. Classical trajectories of strings, and branes, have played a fundamental role in developments within the AdS/CFT correspondence. For example, semiclassical quantization played a central role in the study of BMN \cite{Berenstein:2002jq}, GKP \cite{Gubser:2002tv} and rotating strings  \cite{Frolov:2002av} which can all be understood as classical trajectories of the string with the corresponding fluctuations. The phase space of most mechanical systems is not integrable and thus the role of chaotic classical trajectories needs to be revisited in the context of semiclassical quantization of strings, as originally advocated in  \cite{Zayasa1}. Several applications and examples have recently been provided
\cite{Basu:2011dg,Basu:2012ae,Stepanchuk:2012xi,Basu:2011fw,Basu:2011di,Chervonyi:2013eja}, as will be summarized in section \ref{sec:Analytic}. Certainly much remains to be elucidated, including working out explicitly some entries in the AdS/CFT dictionary under chaotic motion.

A particularly natural class of deformations of field theories are those that are marginal.  Leigh and Strassler obtained powerful results pertaining to the marginal deformations of ${\cal N}=4$ sYM \cite{Leigh:1995ep}. Their work is purely in field theory, where the superpotential of the initial theory is deformed with appropriate exponentials related to  global charges. One of the first attempts to construct the gravitational dual of these exactly marginal deformations was presented in
\cite{Aharony:2002hx}. The problem yielded to an explicit and elegant solution in the particular case of the so-called $\beta$ deformation in \cite{Lunin:2005jy}. The gravity dual background can, in general, be constructed by applying a transformation of the form $S_{\stt} T s_{\gtt} T S^{-1}_{\stt}$ in the original $AdS_5\times S^5$ theory. The inner $T s_{\gtt} T$ transformation refers to a series of T-duality, a shift with parameter $\gtt$ and a T-duality on an appropriate combination of two of $U\prt{1}$ angles of the metric, and produces the real $\beta$ dual background. The additional $S_{\stt}$ deformation denotes an $SL\prt{2,R}$ transformation with a parameter $\stt$ to generate the Lunin-Maldacena (LM) background with complex $\b$. In other words, starting from the $AdS_5\times S^5$ space, a generic appropriate $SL\prt{3,R}$ transformation can be applied to the eight dimensional theory to obtain the complex background.

Having a holographic dual to the $\beta$ deformation of ${\cal N}=4$ sYM opens the gate to answering many questions explicitly. In fact, the real $\beta$-deformed theory has been studied extensively  (see \cite{Zoubos:2010kh} for a review in the context of AdS/CFT integrability) and, in several cases, has been found to resemble closely, although non-trivially, the results of its undeformed parent theory. For example, the expectation values of particular BPS-like Wilson loops \cite{GiataganasWL} remain undeformed. Moreover, a Lax pair for the real $\beta$-deformation was explicitly constructed in  \cite{Frolov:2005dj} by relating the deformed system to the undeformed one, therefore establishing integrability. On the field theory side, the appropriate twist of the integrable structure
was discussed in \cite{Beisert:2005if}.

However, for \emph{complex} $\beta$ the existence of integrability was doubtful from the very beginning. In \cite{Roiban:2003dw}, it was
shown that the 2-scalar field sector enjoys 1-loop integrability, but in \cite{Frolov:2005ty} 2-loop integrability was argued to be problematic. In the 3-scalar sector, already at one-loop the dilatation operator was shown not to map to an integrable hamiltonian \cite{Berenstein:2004ys}. On the gravity side, the intuition \cite{Frolov:2005ty} is that the above-mentioned $S$-dualities necessary to obtain the
complex-$\beta$ background introduce string interactions in intermediate stages, which (as in the study of non-planar effects, see \cite{Kristjansen:2010kg} for a review)
are believed to break integrability. Thus the current consensus is that the complex deformation is not integrable. However a rigorous proof
on the strong-coupling side remains lacking.

%Such was the situation? -2-loop integrability in quivers??- for strings in AdS${}_5\times T^{1,1}$, although integrability arguments were always lacking in that case, until a proof of non-integrability
%was provided in \cite{Basu:2011di}.

Using methods of analytic non-integrability, in this paper we show that a particular complex beta deformation, where the
deformation parameter is taken to be imaginary, {\it is not integrable}. We find a coherent picture of the interplay between
the marginal $\beta$ deformation and integrability. Namely, we analytically prove non-integrability of string motion on the LM
background corresponding to deforming $\Ncal=4$ sYM with an imaginary $\beta$ parameter. We also
perform a numerical analysis of the dynamics of string motion for general complex beta, which shows chaotic-like motion
while a similar analysis for the real-$\beta$ deformation shows no signs of non-integrability.  Our work provides the most
conclusive yet answer to the question of integrability of the marginal deformations of $\Ncal=4$ sYM. %We do so in a closed form.

As part of the proof of  non-integrability in the complex deformed background, we find certain new string solutions. It is worth  remarking that due to the fact that the complex deformed background has an overall conformal factor in the $AdS$ part of the metric and in the complex deformed 5-sphere, as well as a B-field having components along several directions, its system of string equations has major differences compared to that of the real $\beta$ deformed \cite{Frolov:2005ty} and to that of the undeformed theory. % It constrains significantly the string solutions compared to the real $\beta$ deformed theories .
A recent realization of this appeared in the computation of the generalized cusp anomalous dimension \cite{cuspbeta} where in the gravity side for real $\beta$ parameter the system of string equations can be partially mapped to the undeformed $\cN=4$ sYM, while for the complex $\beta$ this is not possible. Another computation pointing out such differences, is the Lax pair construction for the real $\beta$ deformed background,
which is crucially based on a correspondence between the deformed and undeformed string equations \cite{Frolov:2005dj,Alday05tst}, while this mapping can not be made in the complex $\beta$ deformed background.

We should emphasise that our results do not exclude integrability in certain subsectors of string motion. In particular, for the
complex $\beta$ deformation, a certain subsector of $\Ncal=4$ sYM consisting of two holomorphic and one antiholomorphic
scalar has been shown to be integrable at one-loop level \cite{Mansson:2007sh} and, recently, fast spinning string motion in this subsector,
for the purely imaginary deformation,
was argued to be consistent with integrability \cite{Puletti:2011hx}. Our string Ansatz, being more general than that of
\cite{Puletti:2011hx}, is sufficiently generic to exhibit the non-integrability of the background. The search for and classification
of integrable sub-sectors of non-integrable theories remains a very important element in mapping the transition to non-integrability and
we hope that our results will provide new impetus for such work.

The paper is organized as follows. In section \ref{sec:Analytic} we review the essential statements and results in the field of analytic non-integrability of Hamiltonian systems.  Then we review the supergravity background dual to the $\beta$-deformation of ${\cal N}=4 $ supersymmetric Yang-Mills in section \ref{BetaBackground}. We consider the string sigma model in section \ref{String}, and in section \ref{Point} we find an integrable geodesic solution. We analytically prove that the background is non-integrable in section \ref{NonIntegrable}, by finding and studying an extended string solution. In section \ref{Sec:Numerics}, we study numerically the sensitivity of  string motion in the complex $\beta$ deformed background with respect to initial conditions. The behavior we find is in line with our analytic results. We conclude in section \ref{Conclusions} with a summary of our results and pointing out some interesting new directions to explore.

%%%%%%%%%%%%%%%%%%%%%%%%%%%%%%%%%%%%%%%%%%%%%%%%%%%%%%%%%%%%%%%%%%%%%%%%%%%
\section{Analytic non-integrability in Hamiltonian systems}\label{sec:Analytic}
%%%%%%%%%%%%%%%%%%%%%%%%%%%%%%%%%%%%%%%%%%%%%%%%%%%%%%%%%%%%%%%%%%%%%%%%%%%%%%%%
In this section, we briefly review the main statements in the area of analytic\footnote{By analytic we mean meromorphic. A meromorphic function on an open subset D of the complex plane is a function that is holomorphic on all D except a set of isolated points, which are poles of the function.} non-integrability  \cite{Fomenko,Morales-Ruiz,Goriely}. Proving non-integrability of a system of differential equations $\dot{\vec{x}}=\vec{f}(\vec{x})$ is based on the analysis of the variational equation around a particular solution $\bar{x}=\bar{x}(t)$.  The variational equation around $\bar{x}(t)$ is a linear system obtained by linearizing the vector field around $\bar{x}(t)$. If the original nonlinear system admits some first integrals, the variational equation does so as well. It follows that showing that the variational equation does not admit any first integrals within a given class of functions implies that the original nonlinear system is non-integrable. In particular, we are working in the analytic setting where inverting the solution $\bar{x}(t)$ one obtains a (noncompact) Riemann surface $\Gamma$ given by integrating $dt=dw/\dot{\bar{x}}(w)$ with the appropriate limits. By linearizing the system of differential equations around the straight line solution we obtain the {\it Normal Variational Equation } (NVE), which is the component of the linearized system describing the variational normal to the surface $\Gamma$.

The method described here applies to Hamiltonian systems. The relevance to AdS/CFT arises because, in the context of classical strings in the conformal gauge, the Virasoro constraint  provides a Hamiltonian for the systems we consider. The classical string system can be, in certain cases, reduced consistently to a 2-d Hamiltonian system. Given a Hamiltonian system, Ziglin's theorems  \cite{ZiglinI, ZiglinII} connect the existence of a first integral of motion with the monodromy matrices around the straight line solution. In \cite{MR-S,MR-R,MR-R-S}  a major improvement on Ziglin's theory was proposed by introducing techniques of differential Galois theory. It was found that the identity component of the differential Galois group of the variational equations normal to an integrable plane of solutions is Abelian. The calculation of the Galois group is rather intricate, but it can be simplified considerably by applying an algorithm due to Kovacic \cite{Kovacic}, by using the fact that an Abelian identity element in the Galois group is equivalent to finding Liouvillian solutions to the NVE. Liouvillian solutions are those that can be written as combinations of integrals of exponentials, logarithms, rational functions and algebraic expressions. Kovacic's algorithm is an algorithmic implementation of Picard-Vessiot theory for second-order homogeneous linear differential equations with polynomial coefficients,  and provides, in a constructive manner, an answer to the existence of integrability by quadratures.

The above approach to declaring systems non-integrable has been successfully applied to various situations, with some interesting examples including \cite{Morales-Ramis,Mciejewski,ZiglinABC}.
It has also been applied to cosmological models as well as to theories arising in the context of gauge/gravity duality. Signs of chaotic behavior have been found in Schwarzschild black holes \cite{Frolova1,Zayasa1}. More recently the method has been applied to certain classical string configurations relevant in the context of the AdS/CFT correspondence \cite{Basu:2011dg,Basu:2012ae,Stepanchuk:2012xi},
where for example integrability in the Sasaki-Einstein spaces has been ruled out \cite{Basu:2011fw,Basu:2011di}, or has provided evidence that rules out integrability of $N=4$ sYM beyond the planar limit \cite{Chervonyi:2013eja}, although at special large $N$ limits some integrability does appear \cite{Koch:2011hb,Carlson:2011hy}.  

Based on the above, to apply our method we need to consistently reduce the string equations to a
2-d Hamiltonian system, say with variables $\a\prt{\t}$ and $\th\prt{\t}$. Then a solution needs to be found
where an invariant plane is chosen, for instance of the form $\th=c$ and $\dot{\a}=F\prt{\a,\t}$. $F\prt{\a,\t}$ is a generic function of time and $c$ is a constant. Usually, to investigate the integrability of the solution we do not need the full solution of the equations of motion. A full set of variations may be taken, where consistently it should be possible to keep as non-zero only the variation along the normal direction to the invariant plane. We end up with a second order homogeneous differential equation for the normal variations with respect to time:
\be
\ddot{\eta}\prt{t}+F_1\prt{t,\a}\dot{\eta}\prt{t}+F_2\prt{\a,t}\eta\prt{t}=0
\ee
This is the NVE mentioned above. In case the coefficients are not rational, in order to proceed one needs to look for an appropriate redefinition $z=f\prt{\alpha}$ %using the chain rule
%\bea
%\frac{\del}{\del t}=\da \frac{\del z}{\del \a}\frac{\del}{\del z}~,\quad\frac{\del^2 }{\del t^2}=\ddot{\a}\frac{\del z}{\del \a}\frac{\del }{\del z}+\frac{\del^2 z}{\del \a^2} \da^2 \frac{\del}{\del z}+\prt{\frac{\del z}{\del \a}}^2 \da^2 \frac{\del^2}{\del z^2},
%\eea
leading to rational coefficients, where it should be noted that only the equation of motion for $\alpha$ is required, and not the whole solution. In the resulting  second order homogeneous differential equations with rational coefficients we need to look for Liouvillian solutions by applying the Kovacic algorithm. If there exist no such solutions the truncated Hamiltonian system is not integrable and hence the full initial system is not integrable. If we do find such a solution, then the process is inconclusive, since we can not rule out non-integrability of the full system just by finding an integrable truncated Hamiltonian.

%%%%%%%%%%%%%%%%%%%%%%%%%%%%%%%%%%%%%%%%%%%%%%%%%%%%%%%%%%%%%%%%%%%%%%%%%%%%%%%%%%%%%%%%%%%%%%%%%%%%%%%%%%%%%%%5
\section{The Gravity dual Metric of the Complex \texorpdfstring{$\beta$}{beta}-Deformed Theory}\label{BetaBackground}

In the notation of \cite{Frolov:2005ty}, the gravity dual background of the complex $\beta$-deformed $\cN=4$ sYM theory  \cite{Lunin:2005jy} takes the following form:
\bea
\diff s^2&=&R^2 \sqrt{H} \left[ \diff s^2_{\AdS_5}+\sum_{i=1}^3(\diff \rho_i^2+G \rho_i^2\diff \phi_i^2)
+(\gtt^2+\stt^2) G \rho_1^2\rho_2^2\rho_3^2 \left(\sum_{i=1}^3 \diff \phi_i\right)^2 \right]\;,\\
B&=&R^2(\gtt G w_2-12\stt w_1 \diff\psi)\;, \quad \psi=\frac13(\phi_1+\phi_2+\phi_3),\\\nn
w_2&=&\rho_1^2\rho_2^2\diff\phi_1\diff\phi_2-\rho_1^2\rho_3^2\diff\phi_1\diff\phi_3+\rho_2^2\rho_3^2\diff\phi_2\diff\phi_3, \quad \diff w_1=\cos\alpha\sin^3\alpha\sin\theta\cos\theta\diff\alpha\diff\theta,\\\nn
e^\Phi&=&e^{\Phi_0}\sqrt{G} H,
\eea
where the metric, the NS-NS $B$ field and the dilaton have been given. The RR-fields are not presented, since they do not couple directly to the
bosonic part of the classical string action. The functions we have used are the following
\be
G=\frac{1}{1+(\gtt^2+\stt^2) Q}\;,\quad \text{with}\quad  Q=\rho_1^2\rho_2^2+\rho_2^2\rho_3^2+\rho_1^2\rho_3^2,
\ee
and
\be
H=1+\stt^2 Q\;,
\ee
where we have split the real and imaginary parts of $\beta$ as
\be
\beta=\gtt-i \stt\;.
\ee
The Cartesian coordinates $\rho_i$ satisfy $\sum \rho_i^2=1$ and we choose to parametrize them as
\be\la{rhoa1}
\rho_1=\sin\alpha\cos\theta\;,\quad
\rho_2=\sin\alpha\sin\theta\;,\quad
\rho_3=\cos\alpha\;.
\ee
Note that then the functions take a more convenient form
\be
Q=\frac{1}{4}(\sin^2 2\alpha+\sin^4\alpha\sin^2 2\theta)
\ee
and the metric can be written as
\bea\nn
\diff s^2=&&\!\!\!\!\!\!\!\sqrt{H}\prt{-\cosh^2\rho dt^2+d\rho^2}+\sqrt{H} \prt{d\a^2+\sin^2\a d\th^2+G \sum_{i,\prt{ j<k}} \rho_i^2\prt{1+\prt{\gtt^2+\stt^2}\rho_j^2 \rho_k^2}d\phi_i^2 } \\
&&\!\!\!\!+2\sqrt{H} G\prt{\gtt^2+\stt^2}\rho_1^2\rho_2^2\rho_3^2\prt{d\phi_1 d\phi_2 +d\phi_1 d\phi_3+d\phi_2 d\phi_3}~,
\eea
where from the $AdS$ part we have kept only the elements interesting us, and $\rho$ is the radial direction. The indices $i,j,k$ take values from 1 to 3, and the sum $\sum_{i,\prt{ j<k}}$ is used for presentation purposes, and defines a summation in $i$, while $j$ and $k$ take the only remaining allowed values.
%The main difference between the real and complex $\beta$ deformations resides in the form
%\be
%\beta=\tilde{\gamma}-i\tilde{\sigma}~.
%\ee
By taking the smooth limit $\stt\rightarrow 0$ we obtain the usual real $\b$-deformed theory. If the further smooth limit $\gtt\rightarrow 0$ is taken then we are left with the undeformed $\cN=4$ sYM.
For more details on this background we refer the reader to \cite{Frolov:2005ty}.

Without loss of generality we integrate $w_1$ as\footnote{There is clearly an ambiguity at this level since for instance
$w_1=(1/2)\cos \alpha \sin^3 \alpha \sin^2\theta d\alpha$ is also a solution, but it just amounts to a gauge choice that
we can make in a convenient way in order to simplify our equations. See also \cite{Puletti:2011hx} for a relevant discussion. }
\be
w_1=\frac14\sin^4\alpha\cos\theta\sin\theta\diff\theta~.
\ee
For the B-field we thus find
\be
B=R^2 \prt{\gtt G\sum_{i<j}\rho_i\rho_j~\diff\phi_i\wedge\diff\phi_j -\stt\rho_1\rho_2\prt{1-\rho_3^2}\left(\diff\theta\wedge\prt{\diff\phi_1
+\diff\phi_2+\diff\phi_3}\right)}~.
\ee
Notice that in certain cases for presentation and convenience purposes we might use the
$\rho_i$ notation, while in some other their parametrization in angles given in \eq{rhoa1}.

%%%%%%%%%%%%%%%%%%%%%%%%%%%%%%%%%%%%%%%%%%%%%%%%%%%%%%%%%%%%%%%%%%%%%%%%%%%%%%%%%%%%%%%%%%%%%%%%%%%%%%%%%%%%%%%%%%%%%%%%%%%%%%
\section{The String Sigma-model}\label{String}

Let us consider the general ansatz describing a classical string in the deformed dual theory
\be
t=t(\tau),\qquad \rho= \rho\prt{\t}~, \qquad \alpha=\alpha(\tau, \sigma), \qquad  \theta=\theta(\tau, \sigma)~,\qquad \phi_i=\phi_i(\tau, \sigma)~.
\ee
The string is not allowed to have any extension in the space where the field theory lives, while in the internal space the most generic motion is allowed. We are ultimately interested in understanding some consistent particular solutions of this ansatz but it is instructive to start with this level of generality, since we need to verify that the particular string configurations satisfy the full system of equations of motion.

Let us start with the classical string $\sigma$-model action on a general background including an NS-NS $B$-field:
\be
S=-\frac{R^2}2 \int  \diff\tau\frac{\diff\sigma}{2\pi}\left[
\gamma^{\alpha\beta}G_{MN}\p_\alpha X^M\p_\beta X^N-\epsilon^{\alpha\beta}B_{MN} \p_\alpha X^M \p_\beta X^N\right]
\ee
where $\epsilon^{01}=1$ and in conformal gauge $\gamma^{01}=\mathrm{diag}(-1,1)$. We have set $\alpha'=1$ and
extracted the $R^2$ factor from the metric and B-field to emphasise that this is a heavy, classical string.

Substituting our general ansatz above, we arrive at:
\bea\nn
S&=&-\frac{R^2}{4\pi}\int d\s d\t \sqrt{H}\prt{\cosh^2\rho ~\dot{t}^2+\prt{\rho'^2-\dot{\rho}^2}}+ \sqrt{H} \left(\prt{\alpha'^2-\da^2}+\sin^2\a\prt{\th'^2-\dth^2}\right)\\ \nn &&
+\sum_{i}G_{ii} \prt{\phi_i'^2-\dphi_i^2}+
2 \sum_{i,j,\prt{i<j}}G_{ij} \prt{\phi_i'\phi_j'-\dphi_i \dphi_j}-
2\sum_{i,j,\prt{i<j}}B_{ij}\prt{\dphi_i \phi_j'-\phi'_i \dphi_j}\\\la{action1}
&&-2\sum_{i}B_{\theta i}\prt{\dth \phi_i'-\th'\dot{\phi}_i}.
\eea
The general equations of motion for the non-cyclic coordinates $\a$ and $\th$ are given by
\bea\nn
&& \del_{\a,\th}\sqrt{H}\prt{\cosh^2\rho \dot{t}^2+\rho'^2-\dot{\rho}^2}+ \del_{\a,\th}\sqrt{H}\prt{\a'^2-\da^2}+ \del_{\a,\th}\prt{\sqrt{H}\sin^2\a}\prt{\th'^2-\dth^2}\\\nn&&+\del_{\a,\th} G_{ii}\prt{\phi_i'^2-\dphi^2}+
2\sum_{i,j,\prt{i<j}}\del_{\a,\th}G_{ij} \prt{\phi_i'\phi_j'-\dphi_i \dphi_j}-
2\sum_{i,j,\prt{i<j}}\del_{\a,\th} B_{ij}\prt{\dphi_i \phi_j'-\phi'_i \dphi_j}\\&&\nn-2\sum_{i}\del_{\a,\th} B_{\theta i}\prt{\dth \phi_i'-\th'\dphi_i}+\\
&&\begin{cases}& +2\del_0\prt{\sqrt{H}\da}-2\del_1\prt{\sqrt{H}\a'}=0\quad\mbox{for }\a~,\\
&+2\del_0\prt{\sqrt{H}\sin^2\a \dth+B_{\theta i}\phi_i'}-2\del_1\prt{\sqrt{H}\sin^2\a \th'+2 B_{\theta i}\dphi_i}=0\quad\mbox{for } \th~.
\end{cases}
\eea
The equations for the cyclic coordinates $\phi_i$ take the simpler form
\be
\del_0\prt{- \sum_{j} G_{ij}  \dphi_j-
\sum_{j} B_{ij}\phi_j'
+ B_{\theta i}\th'}+\del_1\prt{\sum_{j} G_{ij}  \phi_j'+
\sum_{j} B_{ij}\dphi_j
- B_{\theta i}\dth}=0~.
\ee
Finally the Virasoro constraints read
\bea\nn
&&\sqrt{H}\prt{-\cosh^2\rho \dot{t}^2+\prt{\rho'^2+\dot{\rho}^2}}+ \sqrt{H}\prt{\prt{\a'^2+\da^2}+\sin^2\a\prt{\th'^2+\dth^2}}+\sum_{i}G_{ii} \prt{\phi_i'^2+\dphi_i^2}\\\la{vca1}
&&+
2 \sum_{i,j,\prt{i<j}}G_{ij} \prt{\phi_i'\phi_j'+\dphi_i \dphi_j}=0~,\\\la{vca2}
&&\sqrt{H}\rho'\dot{\rho}+ \sqrt{H}\prt{\a'\da+\sin^2\a\th'\dth}+\sum_{i}G_{ii} \phi_i'\dphi_i+ \sum_{i,j,\prt{i<j}}G_{ij} \prt{\phi_i'\dphi_j+\phi_j'\dphi_i}=0~.
\eea
In the following sections we use the generic equations found here to obtain particular solutions.\footnote{Some special string solutions of the above equations have previously been found in \cite{Frolov:2005ty},\cite{Avramis:2007mv} and \cite{Puletti:2011hx}.} Of course, we
explicitly prove that the reduced string configurations we study satisfy the full system of the equations, as should be the case.

%%%%%%%%%%%%%%%%%%%%%%%%%%%%%%%%%%%%%%%%%%%%%%%%%%%%%%%%%%%%%%%%%%%%%%%%%%%%%%%%%%%%%%%%%%
\section{An Integrable Point-like Solution} \label{Point}

As a warm-up, let us consider a point-like string by taking the following ansatz:
\be
t=t(\tau)~,\quad \rho=\rho\prt{\t}~,\quad \alpha=\alpha(\tau)\;,\quad \theta=\theta(\tau)~,
\ee
where the other embedding coordinates are constants. This ansatz gives a consistent motion for the string, while the motion along all the other coordinates is localized.
The equations of motion for $t$ and $\rho$ give
\bea
&&\dot{t}=\frac{\k}{\sqrt{H}\cosh^2\rho}~,\\
&&2\partial_0\prt{\dot{\rho}\sqrt{H}}+\frac{\k^2}{\sqrt{H}}\partial_\rho\cosh^2\rho=0~,
\eea
where $\k$ is the integration constant. The equations can be solved by turning off the motion in the $\rho$ coordinate and localizing it to the bulk $\rho=0$, resulting in the following equation for $t$
\be\la{tsol}
\dot{t}=\frac{\kappa}{\sqrt{H}}~.
\ee
Then in the non-trivial Virasoro  constraint \eq{vca1} only one term from the $AdS$ part contributes and becomes
\be
-\frac{\k^2}{\sqrt{H}}+\sqrt{H}\dot{\a}^2+\sqrt{H}\sin^2\a\dot{\th}^2=0~.
\ee
The equations of motion for $\a$ and $\theta$ become
\bea
&&\partial_\alpha\sqrt{H} \frac{\k^2}{H}-\del_\a\sqrt{H}\dot{\a}^2-\del_\a\prt{{\sqrt{H}\sin^2\a}}\dot{\th}^2 +2 \del_0\prt{\dot{\a}\sqrt{H}}=0,\\
&&\partial_\th\sqrt{H} \frac{\k^2}{H}-\del_\th\sqrt{H}\dot{\a}^2-\sin^2\a \del_\th{\sqrt{H}}\dot{\th}^2 +2 \del_0\prt{\sqrt{H}\sin^2\a\dot{\th}}=0~.
\eea
This system has an effective Lagrangian\footnote{This is the Lagrangian that leads to the above equations of motion, and
can of course also be derived by appropriately substituting the cyclic coordinate $t$ in the original action (\ref{action1}) following
the Routhian procedure. Note that we will be setting the AdS$_5$ radius $R=1$ from now on.}
\be
2 L_{eff}=\frac{\k^2}{\sqrt{H}}+\sqrt{H}\dot{\a}^2+\sqrt{H}\sin^2\a\dot{\th}^2~,
\ee
reducing to a 2-d particle Hamiltonian system. A solution to the equations of motion may be given by a further localization of the point-like string to the equator of the $S^2$
\be\la{sola1}
\a=\frac{\pi}{2}~,\quad \dot{\th}^2=\frac{\k^2}{H}~.
\ee
We proceed to the study of the normal fluctuations to this solution which is governed by the NVE. Taking the variation along the normal plane
\be
\a=\frac{\pi}{2}+\eta\prt{t}~,
\ee
and keeping up to linear order in $\eta(t)$, the NVE becomes
\be
2 H_0^2 \ddot{\eta}+\frac{\stt^2}{2}H_0\sin 4\th~\dot{\th}\dot{\eta}+\eta\prt{H_0\prt{2 H_0-\frac{\stt^2}{2}\prt{1+\cos^2 2\th}}\dot{\th}^2+\stt^2\prt{1+\cos^2 2\th}\frac{\k^2}{2}}=0~,
\ee
where $H_0= 1+\stt^2\sin^2 2\th/4$. This NVE has non-rational coefficients. To bring it into a desirable form we switch to a new variable $z=\cos\th$
to arrive at
\be\la{point1}
2 \k^2 H_0 \prt{\prt{1-z^2}\eta''\prt{z}- z \eta'\prt{z}+\eta\prt{z}}=0~.
\ee
 To obtain \eq{point1} we have used the equation \eq{sola1}, so that the NVE is calculated for the solution of $\theta\prt{\t}$. We note that the final NVE has no $\stt$ or $\gtt$ dependence and it is undeformed, and therefore it is integrable admitting the Liouvillian solution
\be
\eta\prt{z}=c_1 z+c_2 \sqrt{z^2-1}~.
\ee
It is quite remarkable that the geodesic of the deformed theory, despite clearly being dependent on the deformation parameters, has an NVE that remains undeformed and is the same as in the $\cN=4$ sYM theory. In the next section we note that a special limit of the extended string solution we obtain, is a point-like string which again has no $\stt$ dependent NVE and is integrable, in contrast with the non-integrable extended string.

%%%%%%%%%%%%%%%%%%%%%%%%%%%%%%%%%%%%%%%%%%%%%%%%%%%%%%%%%%%%%%%%%%%%%%%%%%%%%%%%%%%%%%%%%%%%%%%%%%%%%%%%%%%%%%%%%%%%%%%%%%%%%%%%%%%%%%%
\section{Non-Integrable String Solution Extended Along a \texorpdfstring{$U\prt{1}$}{U(1)} Direction}\label{NonIntegrable}

In this section we study extended string solutions in the internal space. We show how integrability is lost when the point like string becomes extended. Moreover we show how the integrability is restored when the deformation complex parameter, $\stt$, goes to zero.% corresponding to an extended string solution of the real $\beta$ deformed space which is known to be integrable [cite, cite].

A generic initial ansatz considered for study for the string motion is
\be \la{ansatz1}
t=t(\tau)~,\quad \alpha=\alpha(\tau)\;,\quad \theta=\theta(\tau)~,\quad
\phi_1=\phi_1(\s),\quad \phi_2=\phi_2(\s),\quad \phi_3=\phi_3(\s)~,
\ee
which corresponds to a string moving along the internal space directions and extended along the $U\prt{1}$ angles.
The solution in the $AdS$ part is still given by the equation \eq{tsol}.

First, we solve the equations for the cyclic coordinates $\phi_i$ by constraining the string extension along them further
\be\la{ansatz2}
\phi_1=\phi_3=0~,\qquad\mbox{and}\qquad\phi_2=m \s~,
\ee
where $m$ is a constant, and by setting the real part of deformation parameter to zero, $\gtt=0$. We are thus looking for
solutions on the imaginary-$\beta$ background.

The remaining equations  take the form
\bea\nn
&\partial_{\a,\th}\sqrt{H}  \frac{\k^2}{H}- \partial_{\a,\th}\sqrt{H}\dot{\a}^2-\partial_{\a,\th}\prt{\sqrt{H}\sin\a^2} \dot{\th}^2+\partial_{\a,\th}\prt{G_{22}}\phi_2'^2-2 \del_{\a,\th}B_{\th\phi_2}\dot{\th}m+\\\la{thext}
&\begin{cases} &+ 2\partial_0\prt{\sqrt{H}\dot{\a}}=0\quad\mbox{eom for }\a~,\\
&+ 2\partial_0\prt{\sqrt{H}\sin\a^2\dot{\th}+B_{\theta i}m}=0\quad\mbox{eom for }\th~,
\end{cases}
\eea
and the Virasoro constraint reads
\be
-\frac{\k^2}{\sqrt{H}}+\sqrt{H}\dot{\a}^2+\sqrt{H}\sin^2\a\dot{\th}^2+G_{22} m^2=0~,
\ee
with $G_{22}=\sqrt{H}G \rho_2^2[1+(\gtt^2+\stt^2)\rho_1^2\rho_3^2]$.
The effective Lagrangian of the system then reads
\be
2L_{eff}=\frac{\k^2}{\sqrt{H}}+\sqrt{H}\dot{\a}^2+\sqrt{H}\sin^2\a\dot{\th}^2 -G_{22} m^2+2 B_{\th\phi_2}\dot{\th} m~,
\ee
and the relevant Hamiltonian
\be\la{hami1}
2{\cal H}=-\frac{\k^2}{\sqrt{H}}+\sqrt{H}\dot{\a}^2+\sqrt{H}\sin^2\a\dot{\th}^2 +G_{22} m^2~,
\ee
is constrained to zero. The solution $\theta=0$ defines an invariant plane, where the equation of motion for the angle $\theta$ is satisfied trivially and
the equation of motion for $\a$ after some manipulation can be written as a total integral
\be
\del_0\prt{-\frac{\k^2}{\sqrt{H}}+\sqrt{H}\dot{\a}^2}=0,
\ee
becoming identical to the Virasoro constraint
\be
\dot{\a}^2=\frac{\k^2}{H}~.
\ee
Varying the equation of motion of $\theta$ \eq{thext}, by setting $\theta=0+\eta\prt{t}$ and keeping up to first order in $\eta(t)$,
we obtain the following NVE
\be
\ddot{\eta}+\prt{2\cot\a+\frac{\stt^2 \sin4\a}{4H_0} } \frac{\k}{\sqrt{H_0}}\dot{\eta}+\prt{m-\frac{2 \stt \k\sin2\a }{H_0}}m\eta=0~,
\ee
where $H_0= 1+\stt^2 \sin^2 2\a/4$. To obtain an NVE with rational coefficients we perform a change of variables
$z=\tan\a$ which gives the following equation
\bea
&&\eta\prt{z}''+\frac{2 }{z}\eta\prt{z}'+\frac{m \left[m \left(\left(\stt^2+2\right) z^2+z^4+1\right)-4 \k \stt z \left(z^2+1\right)\right]}{\k^2 \left(z^2+1\right)^4}\eta\prt{z}=0~.
\eea
This NVE with rational coefficients does not admit any Liouvillian solutions for generic values of the parameters. More precisely the solutions are in terms of the Heun double confluent functions and the system is not integrable.

By taking the limit $m\rightarrow 0$, we localize the string and we end up with a geodesic which differs from the one in the previous section. The resulting NVE does not depend on $\stt$, so it is undeformed, and as a result the differential equation has Liouvillian solutions. By taking the smooth limit $\stt\rightarrow 0$, the extended string solution (and the corresponding NVE) becomes an equation in the undeformed $AdS_5\times S^5$ space and the integrability of the Hamiltonian system is restored as expected, since it gives Liouvillian solutions.

%%%%%%%%%%%%%%%%%%%%%%%%%%%%%%%%%%%%%%%%%%%%%%%%%%%%%%%%%%%%%%%%%%%%%%%%%%%%%%%%%%%%%%%%%%%%%%%%%%%%%%%%%5%%%%%%%%%%%%%%%%%%
\section{Explicit numerical analysis}\label{Sec:Numerics}

In this section we analyze the string equations of motion numerically. The standard analysis is best organized in the Hamiltonian formulation. The aim is to explore its behavior as a dynamical system, in particular its
Poincar\'e sections and the behavior of phase-space trajectories as the deformation parameters are varied.

Starting from the rather general sigma model action \eq{action1} and using the string configuration described by \eq{ansatz1}, \eq{ansatz2}, we rewrite the Hamiltonian \eq{hami1} in terms of the conjugate momenta
\be
\begin{split}
\Hcal=-\frac{\k^2}{2\sqrt{H}}+
\frac{p_\alpha^2}{2\sqrt{H}}+\frac{p_\theta^2}{2\sqrt{H}\sin^2\alpha}
-\frac{B_{\theta\phi_2} m}{\sqrt{H}\sin^2\alpha} p_\theta+ \frac{1}{2}G_{22} m^2+\frac{ B_{\theta\phi_2}^2 m^2}{2\sqrt{H}\sin^2\alpha}~.
\end{split}
\ee
The Hamiltonian is fixed to zero by the Virasoro constraint. To study it numerically, we fix the winding parameter $m$ of the string and the constant $\k$. To study the Poincar\'e sections we further fix $\gtt$ to a non-zero value and vary $\stt$.  Increasing the $\stt$ parameter leads to a destruction of the Kolmogorov-Arnold-Moser (KAM) tori which is a typical indicator of chaotic behavior. More specifically, in Figure 1
we consider three initial conditions which are very close to each other in phase space and track their evolution for a fixed value of the Hamiltonian, drawing a point in the $(\theta,p_\theta)=(q_2,p_2)$ plane each time that the trajectory crosses this plane.\footnote{All the plots in this section have been created with the {\it Maple} {\tt poincare} package.} As it can be seen in the plot, for very small values of the complex deformation parameter $\stt =0.001$ the  three originally nearby trajectories remain close for the whole evolution. In the case $\stt=2.0$ we start noticing quasi-periodic structure and finally for $\stt=10$ the initially nearby points get scattered all over the plane, signalling chaotic behavior, that is, strong sensitivity towards the choice of initial conditions.
\begin{figure}
\vspace{-1.5cm}
\begin{center}
\resizebox{1.7in}{!}{\includegraphics{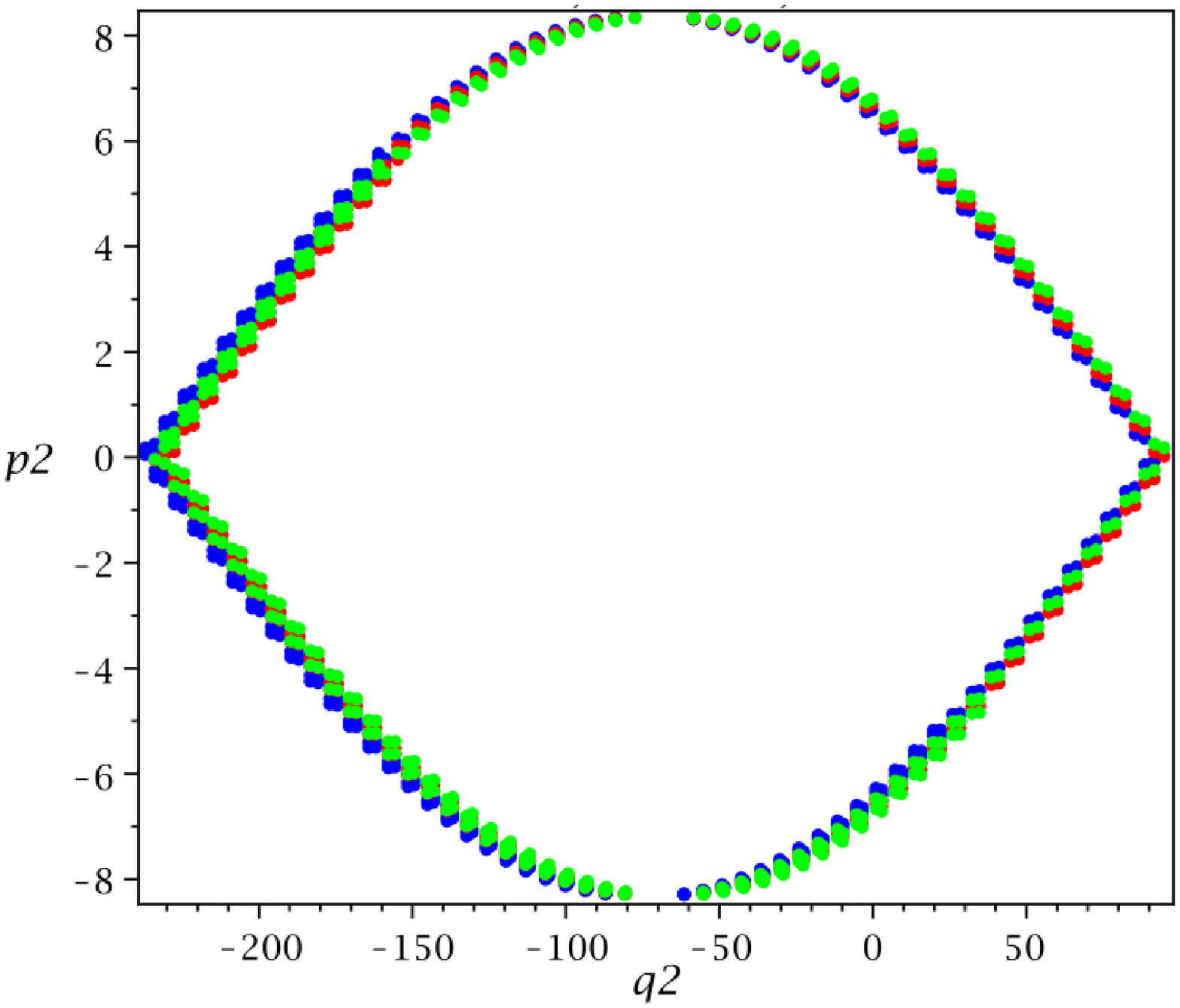}} \,\,\,\,
\resizebox{1.7in}{!}{\includegraphics{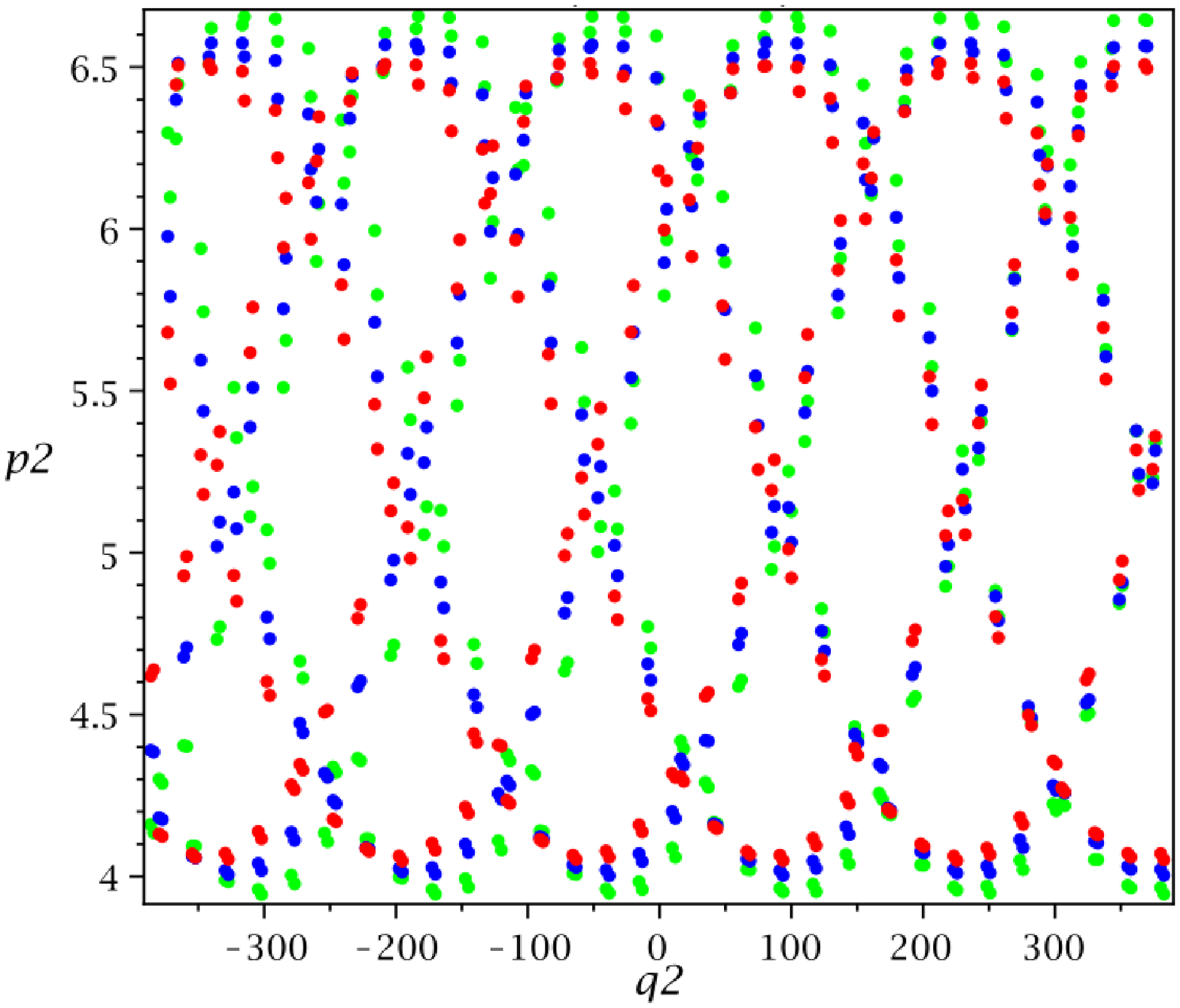}}
\resizebox{1.96in}{!}{\includegraphics{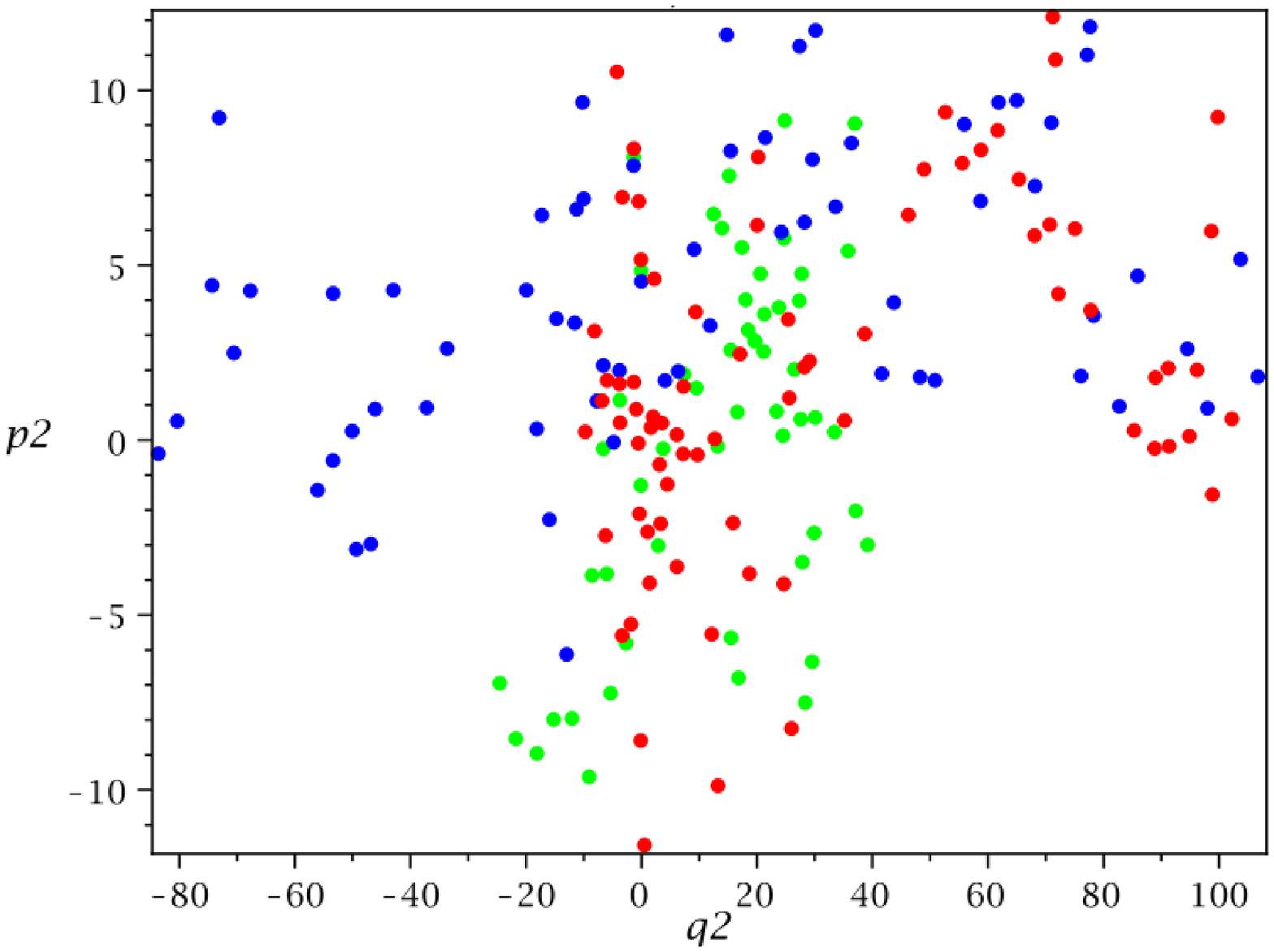}}\vspace{-.2cm}
\caption{\label{Fig.PS} Three Poincar\'e sections for $\gtt=1$ and $\tilde{\sigma}=0.001$ (left panel) $\tilde{\sigma}=2.0$ (center panel) and $\tilde{\sigma}=10$ (right panel). The plots are for $m=2$ and $\kappa=10$.}
\end{center}\vspace{-.8cm}
\begin{center}\hspace{-.8cm}
\includegraphics[width=5.8cm]{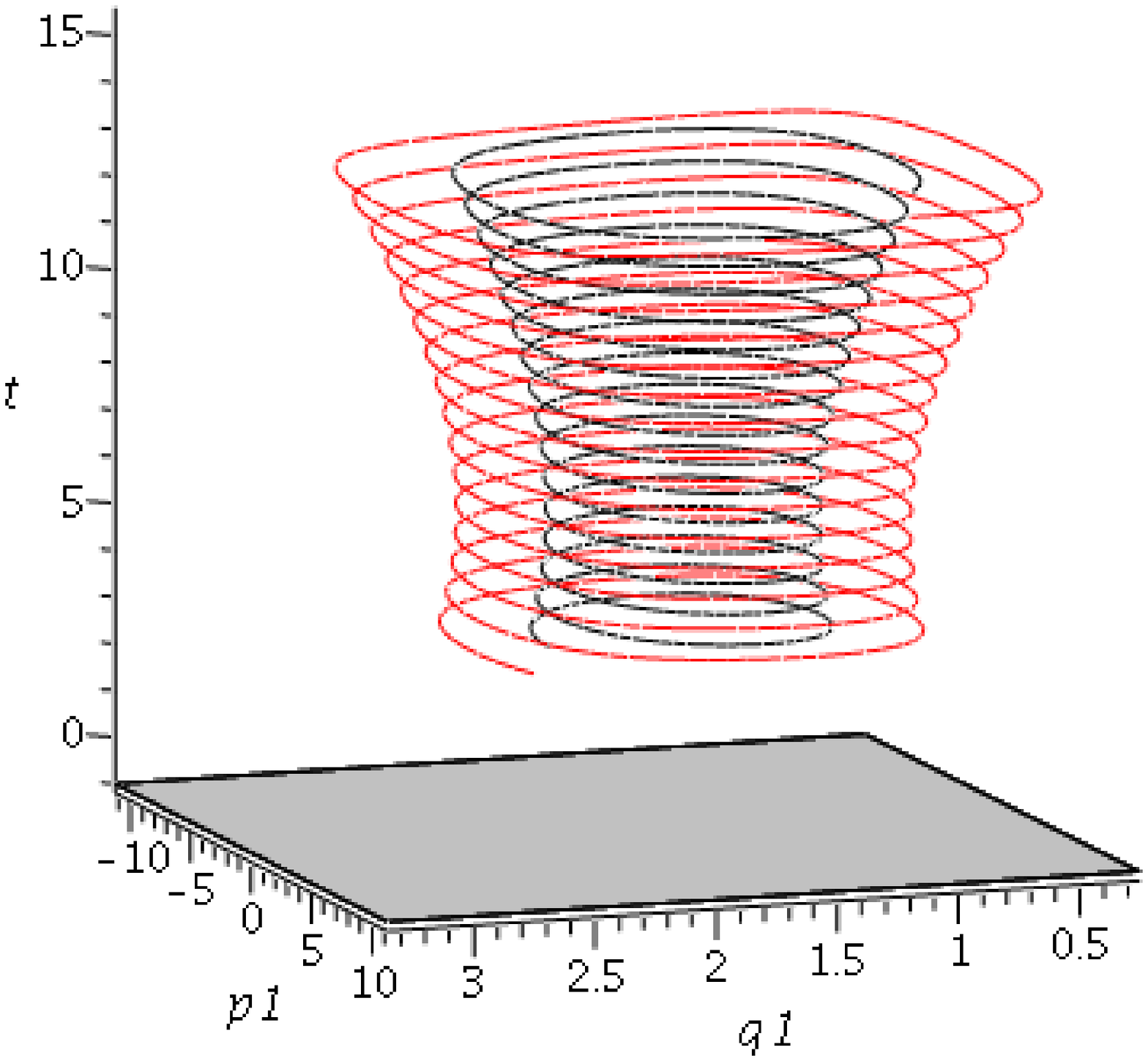} \hspace{-.7cm}
\includegraphics[width=5.8cm]{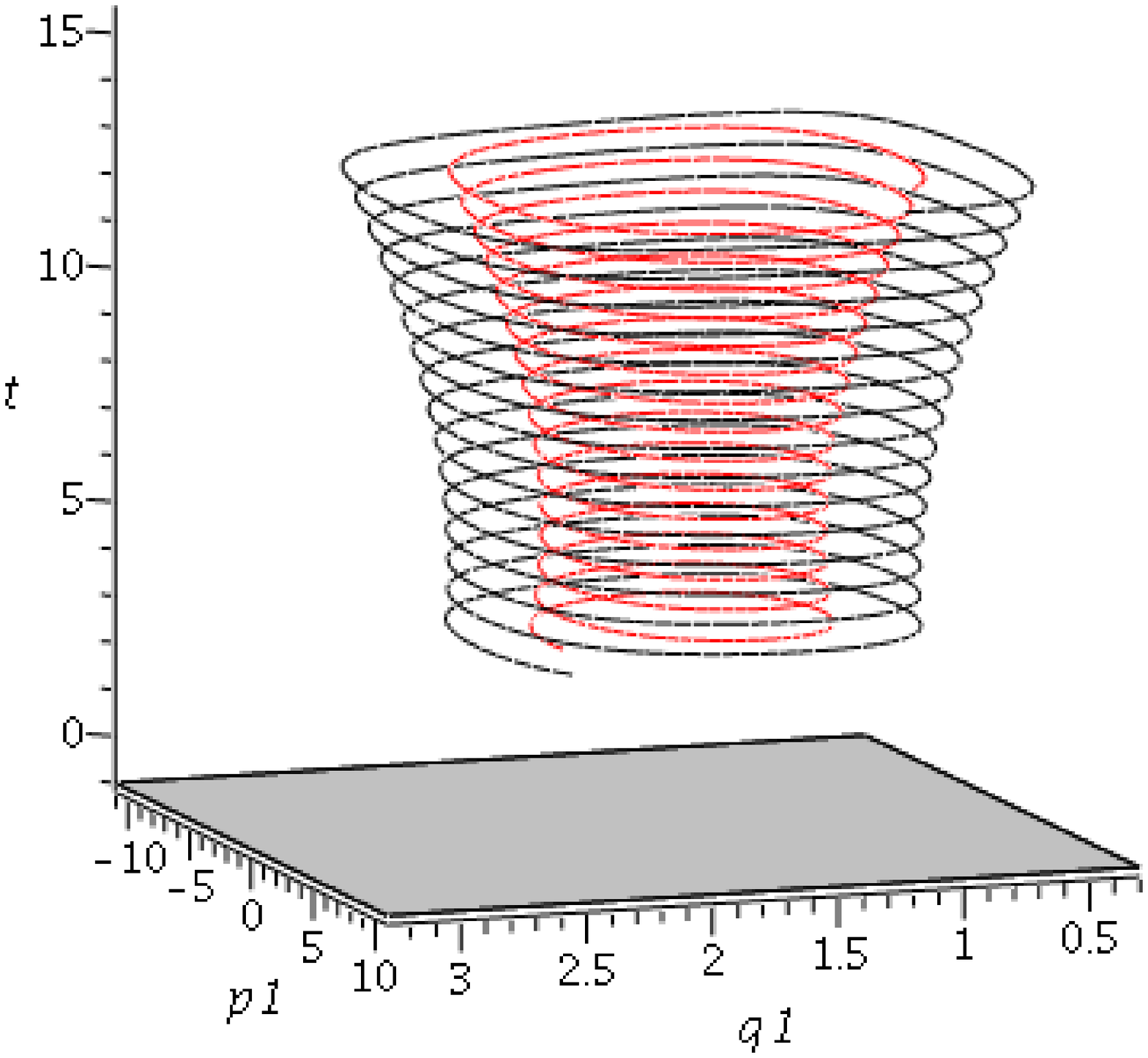} \hspace{-.7cm}
\includegraphics[width=5.8cm]{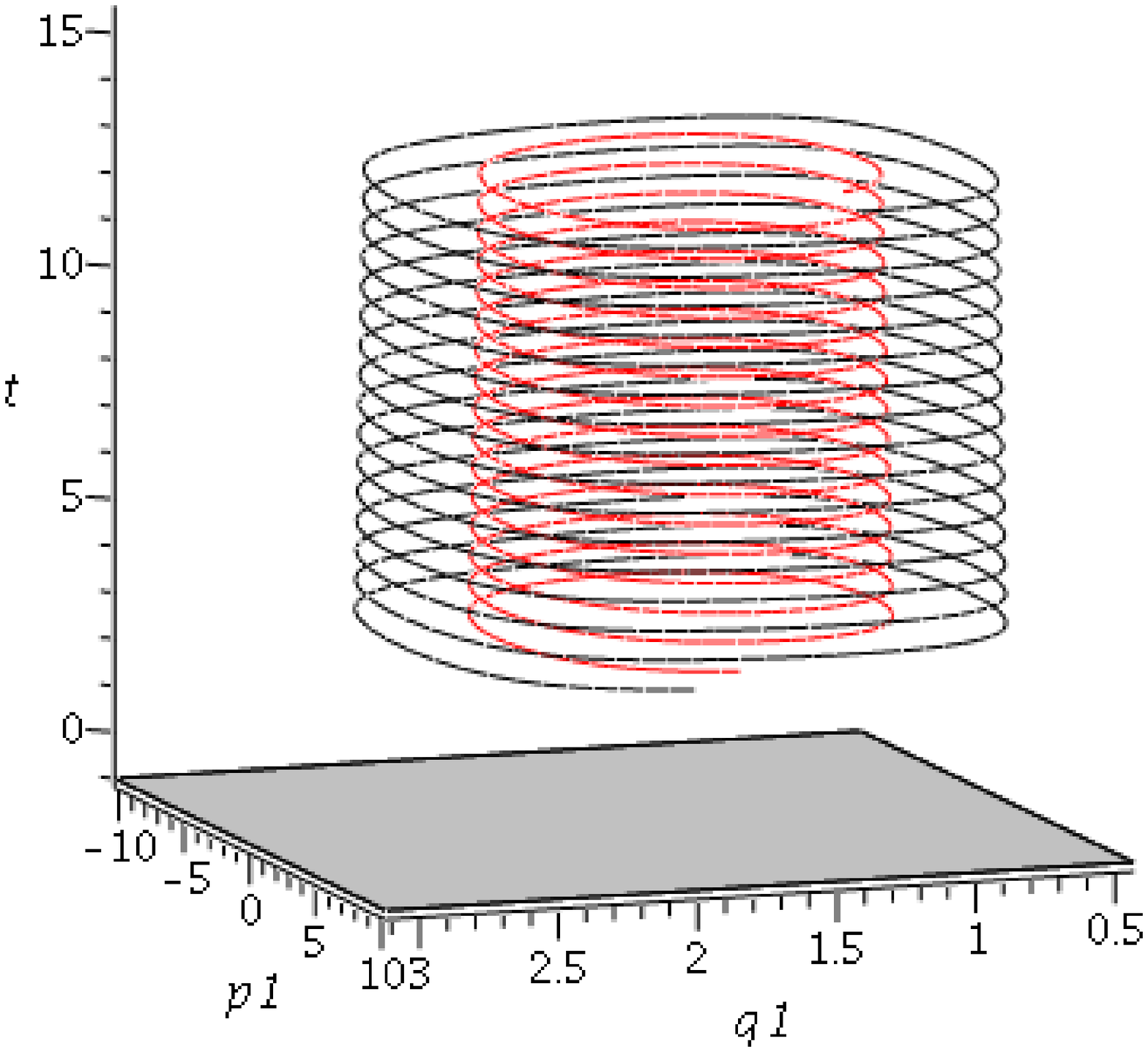}\\\vspace{-.3cm}\hspace{-.7cm}
\includegraphics[width=5.8cm]{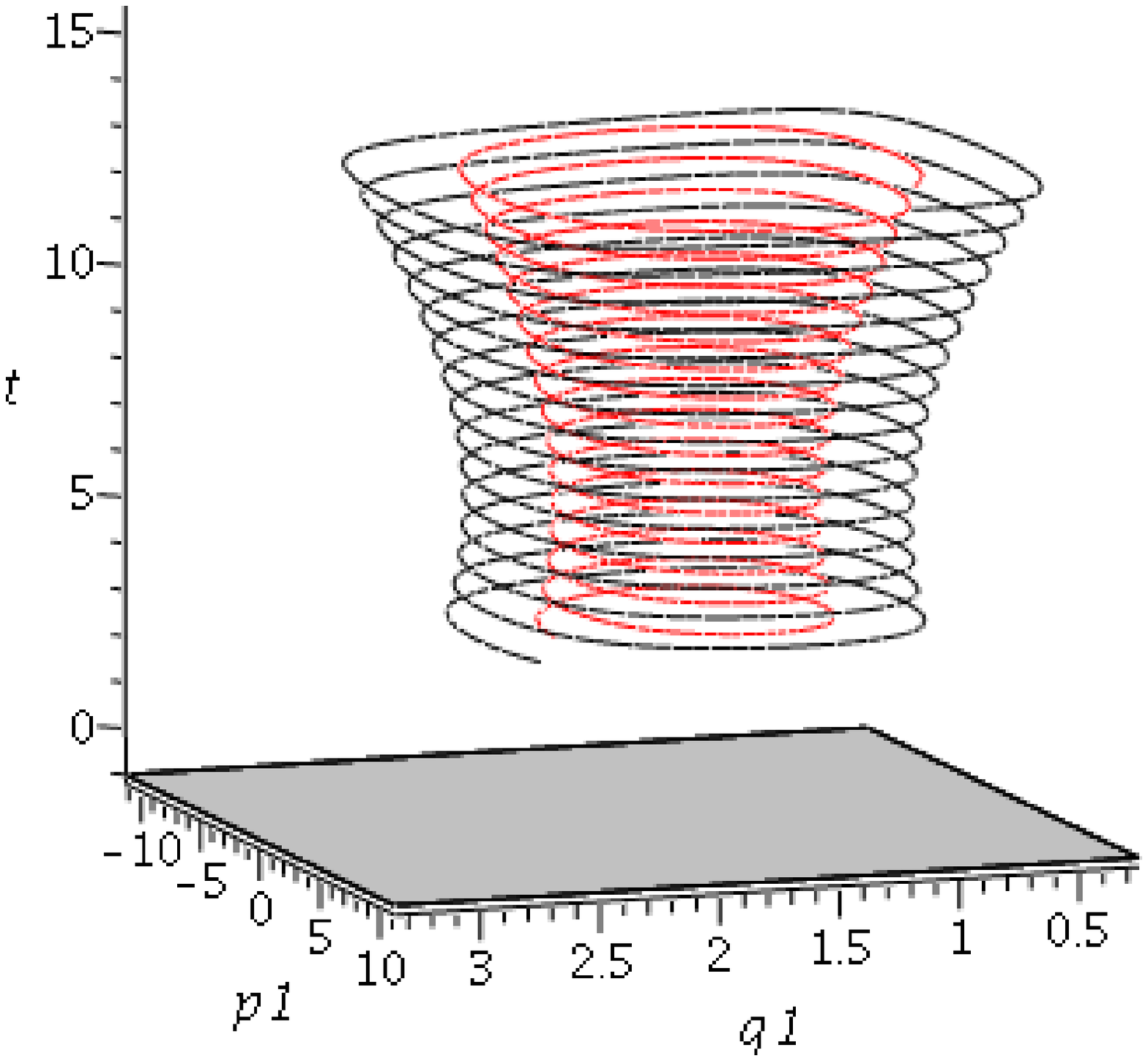} \hspace{-.7cm}
\includegraphics[width=5.8cm]{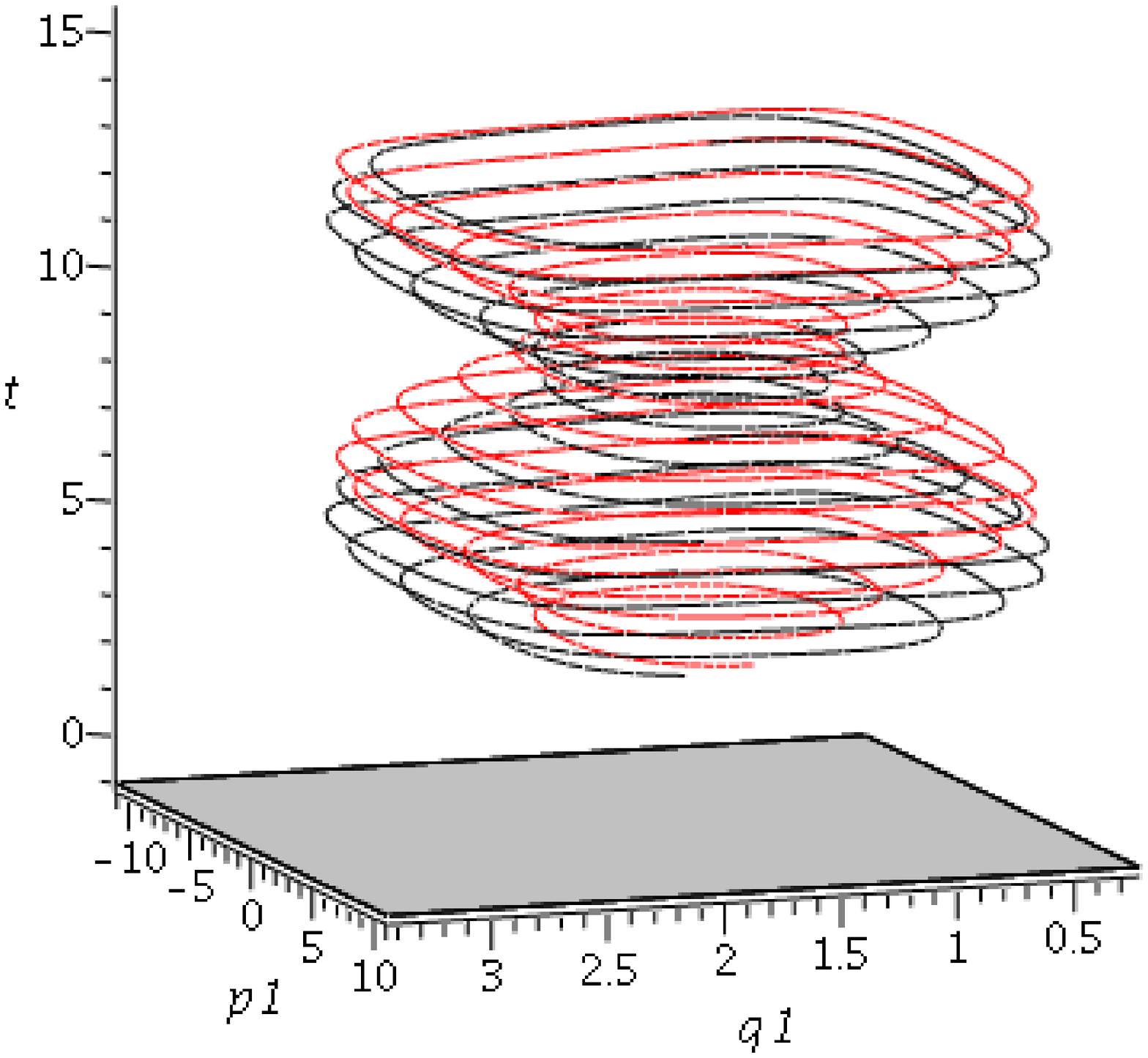}\hspace{-.7cm}
\includegraphics[width=5.8cm]{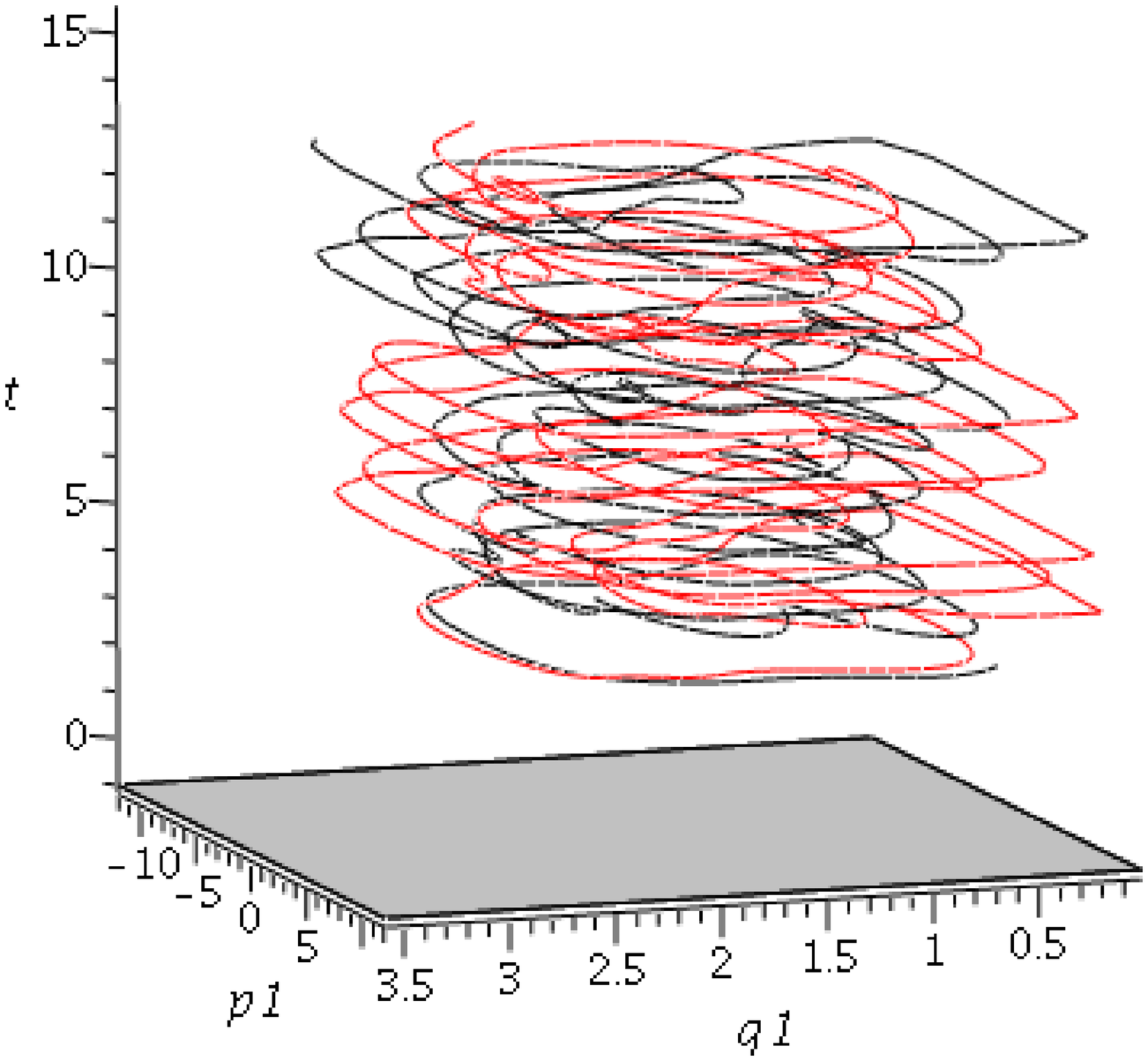}\vspace{-.8cm}
\end{center}
\caption{The top three plots show the phase space trajectories for $\tilde{\sigma}=0$ and  $\tilde{\gamma}=0.01$ (left), $\tilde{\gamma}=1$ (center) and $\tilde{\gamma}=100$ (right). In the bottom three plots, the phase space trajectories for $\tilde{\gamma}=0$ and  $\tilde{\sigma}=0.001$ (left), $\tilde{\sigma}=1$ (center) and $\tilde{\sigma}=10$ (right) are displayed. Each plot is for two nearby initial conditions (black and red trajectory). All plots are for $m=2$ and $\k=10$, with the energy constrained to $0$.  In the top three plots the trajectories remain regular and diverge very slowly from each other, as expected by integrability of the $\tilde{\gamma}$ deformation.
In the bottom three plots, as $\tilde{\sigma}$ increases the trajectories become more irregular and quickly diverge from each other. Note that $(q_1,p_1):=(\alpha,p_\alpha)$.
}
\label{plot3dimaginary}\vspace{-2cm}
\end{figure}

It is also instructive to construct 3-dimensional plots of the phase space trajectories of our system, and compare the $\tilde{\gamma}$ deformation to the $\tilde{\sigma}$ deformation as the deformation parameters increase. Such trajectories are plotted in Figure 2. While the integrability of the $\tilde{\gamma}$ deformation manifests itself in orderly and smooth trajectories, with little sensitivity to initial conditions, in the $\tilde{\sigma}$ deformation the trajectories quickly become irregular as $\tilde{\sigma}$
increases, and show a strong dependence on the initial conditions. This is a signal of chaotic behavior for the complex beta deformation, which directly implies non-integrability. The plots displayed here are for special values ($m=2,\kappa=10$), but we have checked that
very similar behaviour arises for other choices. We have also checked that the chaotic behaviour persists for general complex
$\beta$.

 The results of this section provide strong evidence for the non-integrability of the complex $\beta$ background and
are completely in line with our analytical results in section \ref{NonIntegrable}. Furthermore, since we are not
constrained by the need to find an analytic solution, we can easily check that non-integrable behaviour is present for
general complex $\beta$.

%%%%%%%%%%%%%%%%%%%%%%%%%%%%%%%%%%%%%%%%%%%%%%%%%%%%%%%%%%%%%%%%%%%%%%%%%%%%%%%%%%%%%%%%%%%%%%%%%%%%%%%%%5%%%%%%%%%%%%%%%%%%
\section{Conclusions}\label{Conclusions}

In this paper we have explicitly proved that string motion on the imaginary beta deformed LM background is not integrable. We showed that there exists at least one consistent truncation of the sigma model equations of motion of the imaginary beta deformed theory, that is not an integrable system of ordinary second-order differential equations and this is enough to prove non-integrability of the given sigma model. We arrived at this conclusion using analytical modern methods of Hamiltonian dynamical systems, which we described in detail.  To further support our analytic study we explored numerically the motion of the system and found  it to be chaotic. The chaotic motion arises due to some appropriate noise applied to usually well behaved solutions, resulting to a motion that is highly sensitive to to initial conditions. Non-integrability does not, necessarily, imply chaos. However, the appearance of chaos is evidence of the breakdown of integrability. In our study the chaotic motion of the strings is evident in the Poincar\'e sections and also in the phase space trajectories which show strong sensitivity to the initial conditions. Note that our analytical computations are restricted to the imaginary beta deformation, where it was possible to find an exact solution for string motion. However,
our numerical results apply to the general complex case as well. We thus fully expect that a generalization of our analytical approach to
the generic case of complex $\beta$ will be able to show non-integrability there as well.

Our work motivates a number of further questions.  The interplay between marginal deformations and integrability is one area that should be extensively reconsidered in the context of the AdS/CFT correspondence. We have taken a first step in this paper by addressing the $\beta$ deformation. There is another marginal deformation, the $h$-deformation  \cite{Leigh:1995ep}, which would be interesting to discuss in the same framework. In particular, although the expectation is that the general $(\beta,h)$ deformation will not be integrable,
at the one-loop level there do exist special integrable points in the $(\beta,h)$-plane \cite{Bundzik:2005zg,Mansson:2008xv} (related
to the real-$\beta$ case through appropriate twists in the algebraic structure) and it would
be interesting to check whether string motion on the dual background supports this. Unfortunately, the supergravity background for the $h$
deformation is only known as an expansion to third order in the deformation parameter \cite{Aharony:2002hx,Kulaxizi:2006zc}, but the
analytic non-integrability approach might still be applicable to this order and shed more light on the integrability of this background.

Another interesting question that permeates many areas concerns the quantum implications of classical chaos. The string trajectories that we are discussing should lead, at the quantum level, to excitations of very heavy stringy states. These states should then be identified with operators with large quantum numbers on the field theory side. A similar analysis was performed recently in the nonconformal case, where it was shown that the corresponding hadrons have a Wigner distribution similar to that of realistic hadrons \cite{PandoZayas:2012ig,Basu:2013uva}. Given the generality of the results in quantum chaos, it is sensible to speculate about the existence of a quite universal sector %?(much like in the pp-wave limit) elaborate more maybe..
where many field theories admitting gravity duals will have a similar spectrum of operators dual to highly excited strings.

In all cases that we considered, taking the point-like limit of our string $\sigma$-model led to integrable geodesic motion. We
are not aware of general results on geodesic motion for the complex-deformed LM background, however the integrability or lack thereof
of geodesics on several backgrounds arising in string theory was recently investigated in \cite{Chervonyi:2013eja} and it would
be worthwhile to apply those methods to the LM background as well.

Finally, marginal deformations  can be applied to a wide range of field theories with gravity duals. It would be interesting to investigate the deformed gravity duals of these backgrounds on a wider swath of phase space, including non-integrable trajectories. The prototypical  background is furnished by Sasaki-Einstein  manifolds: $AdS_5\times Y^{p,q}$ \cite{gauntletts2s3}, where the method applied here can be used by generalizing the string solutions obtained in  \cite{GiataganasmSE}. The un-deformed backgrounds are already known to be non-integrable \cite{Basu:2011fw,Basu:2011di}. The hope is, rather, that as in the case of the pp-wave limit \cite{Berenstein:2002jq}, even for these non-integrable backgrounds there is a universal sector that emerges corresponding to classically chaotic strings.

In conclusion, the methods of analytic non-integrability provide a useful
and novel perspective on integrability in the AdS/CFT correspondence and
we believe that further work along the lines we have discussed will be of
great help in mapping the boundaries of integrability and understanding the
mechanisms that lead to its breaking.

\textbf{Acknowledgements:}
% Reference

We are thankful to P. Basu, C.-S. Chu, O. Lunin, J. Russo, K. Sfetsos and A. Tseytlin for useful conversations and comments.  The research of D.G is implemented under the ``ARISTEIA'' action of the ``operational programme education and lifelong learning'' and is co-funded by the European Social Fund (ESF) and National Resources. This work is  partially supported by the Department of Energy under grant
DE-FG02-95ER40899 to the University of Michigan.

%\startappendix
%c
%\Appendix{--}


\begin{thebibliography}{99}
%\providecommand{\href}[2]{#2}\begingroup\raggedright\begin{thebibliography}{10}

%\cite{Maldacena:1997re}
\bibitem{Maldacena:1997re}
  J.~M.~Maldacena,
  ``The large N limit of superconformal field theories and supergravity,''
  Adv.\ Theor.\ Math.\ Phys.\  {\bf 2} (1998) 231
  [Int.\ J.\ Theor.\ Phys.\  {\bf 38} (1999) 1113]
  [arXiv:hep-th/9711200].
  %%CITATION = IJTPB,38,1113;%%

%\cite{Witten:1998qj}
\bibitem{Witten:1998qj}
  E.~Witten,
  ``Anti-de Sitter space and holography,''
  Adv.\ Theor.\ Math.\ Phys.\  {\bf 2} (1998) 253
  [arXiv:hep-th/9802150].
  %%CITATION = 00203,2,253;%%

%\cite{Gubser:1998bc}
\bibitem{Gubser:1998bc}
  S.~S.~Gubser, I.~R.~Klebanov and A.~M.~Polyakov,
  ``Gauge theory correlators from non-critical string theory,''
  Phys.\ Lett.\  B {\bf 428} (1998) 105
  [arXiv:hep-th/9802109].
  %%CITATION = PHLTA,B428,105;%%

%\cite{Beisert:2010jr}
\bibitem{Beisert:2010jr}
  N.~Beisert {\it et al.},
 ``Review of AdS/CFT Integrability: An Overview,''
  Lett.\ Math.\ Phys.\  {\bf 99} (2012) 3
  [arXiv:1012.3982 [hep-th]].
  %%CITATION = ARXIV:1012.3982;%%
  %289 citations counted in INSPIRE as of 14 Oct 2013


%\cite{Berenstein:2002jq}
\bibitem{Berenstein:2002jq}
  D.~E.~Berenstein, J.~M.~Maldacena and H.~S.~Nastase,
  ``Strings in flat space and pp waves from N=4 superYang-Mills,''
  JHEP {\bf 0204} (2002) 013
  [hep-th/0202021].
  %%CITATION = HEP-TH/0202021;%%
  %1388 citations counted in INSPIRE as of 06 Nov 2013

%\cite{Gubser:2002tv}
\bibitem{Gubser:2002tv}
  S.~S.~Gubser, I.~R.~Klebanov and A.~M.~Polyakov,
  ``A Semiclassical limit of the gauge / string correspondence,''
  Nucl.\ Phys.\ B {\bf 636} (2002) 99
  [hep-th/0204051].
  %%CITATION = HEP-TH/0204051;%%
  %729 citations counted in INSPIRE as of 06 Nov 2013


%\cite{Frolov:2002av}
\bibitem{Frolov:2002av}
  S.~Frolov and A.~A.~Tseytlin,
  ``Semiclassical quantization of rotating superstring in AdS(5) x S**5,''
  JHEP {\bf 0206} (2002) 007
  [hep-th/0204226].
  %%CITATION = HEP-TH/0204226;%%
  %485 citations counted in INSPIRE as of 06 Nov 2013


\bibitem{Zayasa1}
  L.~A.~Pando Zayas and C.~A.~Terrero-Escalante,
  ``Chaos in the Gauge / Gravity Correspondence,''
  JHEP {\bf 1009} (2010) 094
  [arXiv:1007.0277 [hep-th]].
  %%CITATION = ARXIV:1007.0277;%%
  %7 citations counted in INSPIRE as of 15 Aug 2013





%\cite{Basu:2011dg}
\bibitem{Basu:2011dg}
  P.~Basu, D.~Das and A.~Ghosh,
  ``Integrability Lost,''
  Phys.\ Lett.\ B {\bf 699} (2011) 388
  [arXiv:1103.4101 [hep-th]].
  %%CITATION = ARXIV:1103.4101;%%
  %7 citations counted in INSPIRE as of 14 Oct 2013


%\cite{Basu:2012ae}
\bibitem{Basu:2012ae}
  P.~Basu, D.~Das, A.~Ghosh and L.~A.~Pando Zayas,
  ``Chaos around Holographic Regge Trajectories,''
  JHEP {\bf 1205} (2012) 077
  [arXiv:1201.5634 [hep-th]].
  %%CITATION = ARXIV:1201.5634;%%
  %4 citations counted in INSPIRE as of 20 Oct 2013


%\cite{Stepanchuk:2012xi}
\bibitem{Stepanchuk:2012xi}
  A.~Stepanchuk and A.~A.~Tseytlin,
  ``On (non)integrability of classical strings in p-brane backgrounds,''
  J.\ Phys.\ A {\bf 46} (2013) 125401
  [arXiv:1211.3727 [hep-th]].
  %%CITATION = ARXIV:1211.3727;%%
  %1 citations counted in INSPIRE as of 14 Oct 2013

  %\cite{Basu:2011fw}
\bibitem{Basu:2011fw}
  P.~Basu and L.~A.~Pando Zayas,
  ``Analytic Non-integrability in String Theory,''
  Phys.\ Rev.\ D {\bf 84} (2011) 046006
  [arXiv:1105.2540 [hep-th]].
  %%CITATION = ARXIV:1105.2540;%%
  %3 citations counted in INSPIRE as of 14 Oct 2013


%\cite{Basu:2011di}
\bibitem{Basu:2011di}
  P.~Basu and L.~A.~Pando Zayas,
  ``Chaos Rules out Integrability of Strings in AdS$_5 \times T^{1,1}$,''
  Phys.\ Lett.\ B {\bf 700} (2011) 243
  [arXiv:1103.4107 [hep-th]].
  %%CITATION = ARXIV:1103.4107;%%
  %11 citations counted in INSPIRE as of 17 Oct 2013


\bibitem{Chervonyi:2013eja}
  Y.~Chervonyi and O.~Lunin,
  ``(Non)-Integrability of Geodesics in D-brane Backgrounds,''
  arXiv:1311.1521 [hep-th].
  %%CITATION = ARXIV:1311.1521;%%


%\cite{Leigh:1995ep}
\bibitem{Leigh:1995ep}
  R.~G.~Leigh and M.~J.~Strassler,
  ``Exactly marginal operators and duality in four-dimensional N=1 supersymmetric gauge theory,''
  Nucl.\ Phys.\ B {\bf 447}, 95 (1995)
  [hep-th/9503121].
  %%CITATION = HEP-TH/9503121;%%
  %452 citations counted in INSPIRE as of 14 Oct 2013


%\cite{Aharony:2002hx}
\bibitem{Aharony:2002hx}
  O.~Aharony, B.~Kol and S.~Yankielowicz,
  ``On exactly marginal deformations of N=4 SYM and type IIB supergravity on AdS(5) x S**5,''
  JHEP {\bf 0206} (2002) 039
  [hep-th/0205090].
  %%CITATION = HEP-TH/0205090;%%
  %56 citations counted in INSPIRE as of 17 Oct 2013


%\cite{Lunin:2005jy}
\bibitem{Lunin:2005jy}
  O.~Lunin and J.~M.~Maldacena,
  ``Deforming field theories with U(1) x U(1) global symmetry and their gravity duals,''
  JHEP {\bf 0505} (2005) 033
  [hep-th/0502086].
  %%CITATION = HEP-TH/0502086;%%
  %320 citations counted in INSPIRE as of 14 Oct 2013


%\cite{Zoubos:2010kh}
\bibitem{Zoubos:2010kh}
  K.~Zoubos,
  ``Review of AdS/CFT Integrability, Chapter IV.2: Deformations, Orbifolds and Open Boundaries,''
  Lett.\ Math.\ Phys.\  {\bf 99} (2012) 375
  [arXiv:1012.3998 [hep-th]].
  %%CITATION = ARXIV:1012.3998;%%
  %40 citations counted in INSPIRE as of 14 Oct 2013

  \bibitem{GiataganasWL}
  C.~S.~Chu and D.~Giataganas,
  ``Near BPS Wilson Loop in beta-deformed Theories,''
  JHEP {\bf 0710} (2007) 108
  [arXiv:0708.0797 [hep-th]].
  %%CITATION = JHEPA,0710,108;%%


%\cite{Frolov:2005dj}
\bibitem{Frolov:2005dj}
  S.~Frolov,
  ``Lax pair for strings in Lunin-Maldacena background,''
  JHEP {\bf 0505} (2005) 069
  [hep-th/0503201].
  %%CITATION = HEP-TH/0503201;%%
  %167 citations counted in INSPIRE as of 17 Oct 2013

%\cite{Beisert:2005if}
\bibitem{Beisert:2005if}
  N.~Beisert and R.~Roiban,
  ``Beauty and the twist: The Bethe ansatz for twisted N=4 SYM,''
  JHEP {\bf 0508} (2005) 039
  [hep-th/0505187].
  %%CITATION = HEP-TH/0505187;%%
  %102 citations counted in INSPIRE as of 11 Nov 2013


%\cite{Roiban:2003dw}
\bibitem{Roiban:2003dw}
  R.~Roiban,
  ``On spin chains and field theories,''
  JHEP {\bf 0409} (2004) 023
  [hep-th/0312218].
  %%CITATION = HEP-TH/0312218;%%
  %67 citations counted in INSPIRE as of 17 Oct 2013


%\cite{Frolov:2005ty}
\bibitem{Frolov:2005ty}
  S.~A.~Frolov, R.~Roiban and A.~A.~Tseytlin,
  ``Gauge-string duality for superconformal deformations of N=4 super Yang-Mills theory,''
  JHEP {\bf 0507} (2005) 045
  [hep-th/0503192].
  %%CITATION = HEP-TH/0503192;%%
  %136 citations counted in INSPIRE as of 17 Oct 2013


%\cite{Berenstein:2004ys}
\bibitem{Berenstein:2004ys}
  D.~Berenstein and S.~A.~Cherkis,
  ``Deformations of N=4 SYM and integrable spin chain models,''
  Nucl.\ Phys.\ B {\bf 702} (2004) 49
  [hep-th/0405215].
  %%CITATION = HEP-TH/0405215;%%
  %98 citations counted in INSPIRE as of 17 Oct 2013

%\cite{Kristjansen:2010kg}
\bibitem{Kristjansen:2010kg}
  C.~Kristjansen,
  ``Review of AdS/CFT Integrability, Chapter IV.1: Aspects of Non-Planarity,''
  Lett.\ Math.\ Phys.\  {\bf 99} (2012) 349
  [arXiv:1012.3997 [hep-th]].
  %%CITATION = ARXIV:1012.3997;%%
  %25 citations counted in INSPIRE as of 21 Oct 2013


\bibitem{cuspbeta}
  G.~Georgiou and D.~Giataganas,
  ``Generalised cusp anomalous dimension in beta-deformed super Yang Mills theory,''
  arXiv:1306.6620 [hep-th].
  %%CITATION = ARXIV:1306.6620;%%

\bibitem{Alday05tst}
  L.~F.~Alday, G.~Arutyunov and S.~Frolov,
  ``Green-Schwarz strings in TsT-transformed backgrounds,''
  JHEP {\bf 0606} (2006) 018
  [arXiv:hep-th/0512253].
  %%CITATION = JHEPA,0606,018;%%

%\cite{Mansson:2007sh}
\bibitem{Mansson:2007sh}
  T.~M\aa nsson,
  ``The Leigh-Strassler Deformation and the Quest for Integrability,''
  JHEP {\bf 0706} (2007) 010
  [hep-th/0703150].
  %%CITATION = HEP-TH/0703150;%%
  %13 citations counted in INSPIRE as of 10 Nov 2013

%\cite{Puletti:2011hx}
\bibitem{Puletti:2011hx}
  V.~G.~M.~Puletti and T.~M\aa nsson,
  ``The dual string sigma-model of the $SU_q(3)$ sector,''
  JHEP {\bf 1201} (2012) 129
  [arXiv:1106.1116 [hep-th]].
  %%CITATION = ARXIV:1106.1116;%%
  %1 citations counted in INSPIRE as of 10 Nov 2013


\bibitem{Goriely}
A. Goriely, ``Integrability and Nonintegrability of Dynamical Systems,'' World Scientific, 2001.

\bibitem{Fomenko}
A. T. Fomenko, `` Integrability and Nonintegrability in Geometry and Mechanics,'' Kluwer Academic Publishers, 1988.

\bibitem{Morales-Ruiz}
 Juan Jos\'e Morales Ruiz, ``Differential Galois theory and non-integrability of Hamiltonian Systems, '' Birkh\"auser, Basel 1999.


\bibitem{ZiglinI}
S.L. Ziglin, ``Branching of solutions and non-existence of first integrals in
Hamiltonian mechanics I,'' Funct. Anal. Appl. 16 (1982), 181-189.

\bibitem{ZiglinII}
S. L. Ziglin, `` Branching of solutions and non-existence of first integrals in Hamiltonian me-
chanics II,'' Funct. Anal. Appl., 17 (1983), pp. 617.


\bibitem{MR-S}
J.J. Morales-Ruiz and C. Sim\'o, ``Picard-Vessiot theory and Ziglin's Theorem,'' J. Differential Equations {\bf 107}  (1994) 140

\bibitem{MR-R-S}
J.J. Morales-Ruiz, J.-P. Ramis and C. Sim\`o.  ``Integrability of
Hamiltonian systems and differential Galois groups of higher variational
equations. Ann. Sci. c. Norm. Supr. (4) {\bf 40} (2007) 845884


\bibitem{MR-R}
J.J. Morales-Ruiz and J. P. Ramis,``Galoisian obstructions to integrability of Hamiltonian Systems I \& II, '' Methods Appl.Anal. {\bf 8}  (2001) 33-111


\bibitem{Kovacic}
J.J. Kovacic, ``An Algorithm for Solving Second Order Linear Homogeneous Differential Equations,'' J. Symb. Comput. {\bf 2} (1986), 3.


\bibitem{Morales-Ramis}
J. J. Morales-Ruiz and  J. P. Ramis, `` A Note on the Non-Integrability of Some Hamiltonian Systems with a Homogeneous Potential, ''
Methods and Applications of Analysis {\bf 8} 1 (2001) 113.

\bibitem{Mciejewski}
A.J Maciejewski and M. Szydlowski, ``Integrability and Non-Integrability of Planar
Hamiltonian Systems of Cosmological Origin,'' J. Nonl. Math. Phys. 2001, V.8, Supplement, 200206 Proceedings: NEEDS99


\bibitem{ZiglinABC}
S.L. Ziglin, ``An Analytic Proof of the Nonintegrability of the ABC-flow for $A=B=C$,'' Funct. Anal. Appl.,{\bf 37} 3 (2003) 225.

\bibitem{Frolova1}
  A.~V.~Frolov and A.~L.~Larsen,
  ``Chaotic scattering and capture of strings by black hole,''
  Class.\ Quant.\ Grav.\  {\bf 16} (1999) 3717
  [gr-qc/9908039].
  %%CITATION = GR-QC/9908039;%%
  %14 citations counted in INSPIRE as of 15 Aug 2013


\bibitem{Koch:2011hb}
  R.~de Mello Koch, M.~Dessein, D.~Giataganas and C.~Mathwin,
  ``Giant Graviton Oscillators,''
  JHEP {\bf 1110} (2011) 009
  [arXiv:1108.2761 [hep-th]].
  %%CITATION = ARXIV:1108.2761;%%
  %21 citations counted in INSPIRE as of 02 Nov 2013

\bibitem{Carlson:2011hy}
  W.~Carlson, R.~de Mello Koch and H.~Lin,
  ``Nonplanar Integrability,''
  JHEP {\bf 1103} (2011) 105
  [arXiv:1101.5404 [hep-th]].
  %%CITATION = ARXIV:1101.5404;%%
  %25 citations counted in INSPIRE as of 02 Nov 2013

\bibitem{Avramis:2007mv}
  S.~D.~Avramis, K.~Sfetsos and K.~Siampos,
  ``Stability of string configurations dual to quarkonium states in AdS/CFT,''
  Nucl.\ Phys.\ B {\bf 793} (2008) 1
  [arXiv:0706.2655 [hep-th]].
  %%CITATION = ARXIV:0706.2655;%%
  %22 citations counted in INSPIRE as of 10 Nov 2013

%\cite{Bundzik:2005zg}
\bibitem{Bundzik:2005zg}
  D.~Bundzik and T.~M\aa nsson,
  ``The General Leigh-Strassler deformation and integrability,''
  JHEP {\bf 0601} (2006) 116
  [hep-th/0512093].
  %%CITATION = HEP-TH/0512093;%%
  %21 citations counted in INSPIRE as of 10 Nov 2013

%\cite{Mansson:2008xv}
\bibitem{Mansson:2008xv}
  T.~M\aa nsson and K.~Zoubos,
  ``Quantum Symmetries and Marginal Deformations,''
  JHEP {\bf 1010} (2010) 043
  [arXiv:0811.3755 [hep-th]].
  %%CITATION = ARXIV:0811.3755;%%
  %8 citations counted in INSPIRE as of 10 Nov 2013


%\cite{Kulaxizi:2006zc}
\bibitem{Kulaxizi:2006zc}
  M.~Kulaxizi,
  ``Marginal Deformations of N=4 SYM and Open vs. Closed String Parameters,''
  hep-th/0612160.
  %%CITATION = HEP-TH/0612160;%%
  %11 citations counted in INSPIRE as of 10 Nov 2013



%\cite{PandoZayas:2012ig}
\bibitem{PandoZayas:2012ig}
  L.~A.~Pando Zayas and D.~Reichmann,
  ``A String Theory Explanation for Quantum Chaos in the Hadronic Spectrum,''
  JHEP {\bf 1304} (2013) 083
  [arXiv:1209.5902 [hep-th]].
  %%CITATION = ARXIV:1209.5902;%%
  %1 citations counted in INSPIRE as of 14 Oct 2013


%\cite{Basu:2013uva}
\bibitem{Basu:2013uva}
  P.~Basu and A.~Ghosh,
  ``Confining Backgrounds and Quantum Chaos in Holography,''
  arXiv:1304.6348 [hep-th].
  %%CITATION = ARXIV:1304.6348;%%



\bibitem{gauntletts2s3}
  J.~P.~Gauntlett, D.~Martelli, J.~Sparks and D.~Waldram,
  ``Sasaki-Einstein metrics on S(2) x S(3),''
  Adv.\ Theor.\ Math.\ Phys.\  {\bf 8} (2004) 711
  [arXiv:hep-th/0403002].
  %%CITATION = 00203,8,711;%%


\bibitem{GiataganasmSE}
  D.~Giataganas,
  ``Semiclassical strings in marginally deformed toric AdS/CFT,''
  JHEP {\bf 1112} (2011) 051
  [arXiv:1010.1502 [hep-th]].\\
  %%CITATION = ARXIV:1010.1502;%%
  %5 citations counted in INSPIRE as of 09 Nov 2013


\end{thebibliography}
\end{document}